\documentclass[aps,prl,amsmath,amssymb,superscriptaddress,showkeys,showpacs,twocolumn,floatfix]{revtex4-1}

\usepackage{amsmath,amssymb,mathtools}	
\usepackage{graphicx,color,subfigure}
\usepackage{float}
\usepackage{appendix}
\usepackage{hyperref}
\usepackage{booktabs}

\begin{document}

\title{Numerical simulations on First-order phase transition through thermal fluctuation}

\author{ Ligong Bian }\email{lgbycl@cqu.edu.cn}

\affiliation{Department of Physics and Chongqing Key Laboratory for Strongly Coupled Physics, Chongqing University, Chongqing 401331, P. R. China}

\author{Yuefeng Di}\email{diyuefeng@itp.ac.cn}
\affiliation{
CAS Key Laboratory of Theoretical Physics, Institute of Theoretical Physics, 
Chinese Academy of Sciences, Beijing 100190, China}
\affiliation{School of Physical Sciences, University of Chinese Academy of Sciences (UCAS), Beijing 100049, China}

\author{Yongtao Jia}\email{20175962@cqu.edu.cn}
\affiliation{Department of Physics and Chongqing Key Laboratory for Strongly Coupled Physics, Chongqing University, Chongqing 401331, P. R. China}

\author{Yang Li}
\affiliation{
CAS Key Laboratory of Theoretical Physics, Institute of Theoretical Physics, Chinese Academy of Sciences, Beijing 100190, China}
\affiliation{School of Physical Sciences, University of Chinese Academy of Sciences, No. 19A Yuquan Road, Beijing 100049, China}

\author{Kehao Zeng}
\affiliation{Department of Physics and Chongqing Key Laboratory for Strongly Coupled Physics, Chongqing University, Chongqing 401331, P. R. China}

\begin{abstract}
In this Letter, we numerically present the possibility of the first-order phase transition occurring through the thermal fluctuation in the early universe. We find that when the temperature is slightly higher than the mass scale of the background field, the bubble-like field configurations appear proceeded by oscillons, which expand and collide to finish the phase transition. We provide the false vacuum decay rate and the accompanied gravitational waves. We also present the vacuum phase transition comparison of the quantum tunneling case and thermal fluctuation case.   

\end{abstract}

\maketitle

\noindent{\it \bfseries  Introduction.} 
In recent years, the cosmological first-order phase transition (PT)
is of growing interest in particle cosmology community, since it is widely predicted by beyond the Standard Model (SM) of particle physics~\cite{Athron:2023xlk} that can be probed by colliders~\cite{Arkani-Hamed:2015vfh}, and
provides the non-equilibrium condition to generate the baryon asymmetry of the Universe through electroweak baryogenesis~\cite{Morrissey:2012db}. The cosmological first-order PT proceeds through vacuum bubble nucleation, expansion, and collision, and sources 
the shear stresses to generate gravitational waves (GWs)~\cite{Witten:1984rs,Hogan:1986dsh}, which is a primary target of space-based interferometers, such as: Laser Interferometer Space Antenna (LISA)~\cite{LISA:2017pwj}, TianQin~\cite{Luo_2016}, and Taiji~\cite{taiji}.

The vacuum bubble nucleation can occur through quantum tunneling~\cite{Coleman:1977py,Callan:1977pt} and/or thermal fluctuation~\cite{Linde:1981zj} in the false vacuum of the early Universe.  Previous numerical simulations on GW production from the first-order PT usually implement the bubble nucleation process by giving the initial bubble configurations and the bubble nucleation rate~\cite{Li:2023yzq,Di:2020kbw,Zhao:2022cnn,Cutting:2018tjt,Cutting:2019zws,Cutting:2020nla,Cutting:2022zgd,Hindmarsh:2013xza,Hindmarsh:2015qta,Hindmarsh:2017gnf}, where the thermal fluctuation effects are absent. When thermal effects are included, it was shown that long-lived, spherically symmetric, localized configurations, oscillons, emerge and can affect the bubble nucleation process~\cite{Copeland:1995fq,Gleiser:1996jb,Gleiser:1993pt,Gleiser:2004iy,Gleiser:2007ts,Copeland:1995fq}. Recent study further shows that theoretically predicted thermal decay rate matches well with the (1+1)-dimensional numerical simulation where vacuum decay proceeded by oscillon precursor transitioning to bubble nucleation in the thermal background~\cite{Pirvu:2023plk}. 

In this Letter, we have investigated the thermal false vacuum decay through (3+1)-dimensional lattice simulation. We have numerically confirmed the nucleation of vacuum bubbles occurs automatically after the formation of oscillons, rather than being manually placed. For the first time, we have numerically examined the vacuum bubble nucleation theory, studied the relation between the false vacuum decay process and the bubble dynamics during the PT, and gave the GWs generated from bubbles collision.

\noindent{\it \bfseries  The simulation framework.} 
We consider the following action within the framework of a non-expanding spacetime background:
\begin{equation}
S=-\int d^4x[\frac{1}{2}\partial^\mu\phi\partial_\mu\phi+V(\phi)]\;.
\end{equation}
The scalar potential function takes the form of~\cite{Dunne:2005rt,Baacke:2003uw,Giblin:2013kea,Child:2012qg,Giblin:2014qia,Giblin:2013kea,Cutting:2018tjt,Cutting:2020nla} 
\begin{equation}
    V(\phi)=\frac{1}{2}M^2\phi^2+\frac{1}{3}\delta\phi^3+\frac{1}{4}\lambda\phi^4 \;,
\end{equation}
where $\delta=-1.632$, $\lambda=0.5$. We study the bubble nucleation occurs automatically at finite temperature rather than under the bubble nucleation rate and bubble profile by hand as in Refs.~\cite{Cutting:2018tjt,Cutting:2020nla,Giblin:2013kea,Child:2012qg,Giblin:2014qia,Giblin:2013kea,Cutting:2018tjt,Cutting:2020nla}. We employ the open source ${\mathcal CosmoLattice}$~\cite{Figueroa:2020rrl,Figueroa:2021yhd} to implement discretization of the field equations and evolve the system using a second-order leap-frog integration scheme. 
 The phase transition completion time scale $M^{-1}$ and the expansion time $H^{-1}$ of the simulation framework satisfy the valid condition $M^{-1} \ll H^{-1}$. All the coefficients in the potential function have been re-scaled with $M$, as well as the field variables and space-time variables. The space and time interval is chosen as $\delta \tilde{x} = 0.44$ and $\delta \tilde{t}=0.088$, which are resolved well for the bubble dynamics. The initial field amplitude and its time derivative are determined by the thermal distribution parameterized by the initial temperature $T_i$. We perform the lattice simulation with the grid number $N^3=512^3$ at different initial temperatures $T_i = 1.1 \sim 1.4 M$ with the interval $\Delta T  = 0.05 M$. The hubble parameter is given by $H = 0.33 g_*^{1/2}T^2/m_{pl} $ with the effective number of relativistic degrees of freedom $g_* = 106.75$ and the reduced Planck mass $m_{pl}$.

\begin{figure}[!htp]
    \centering    
    \includegraphics[width=0.225\textwidth]{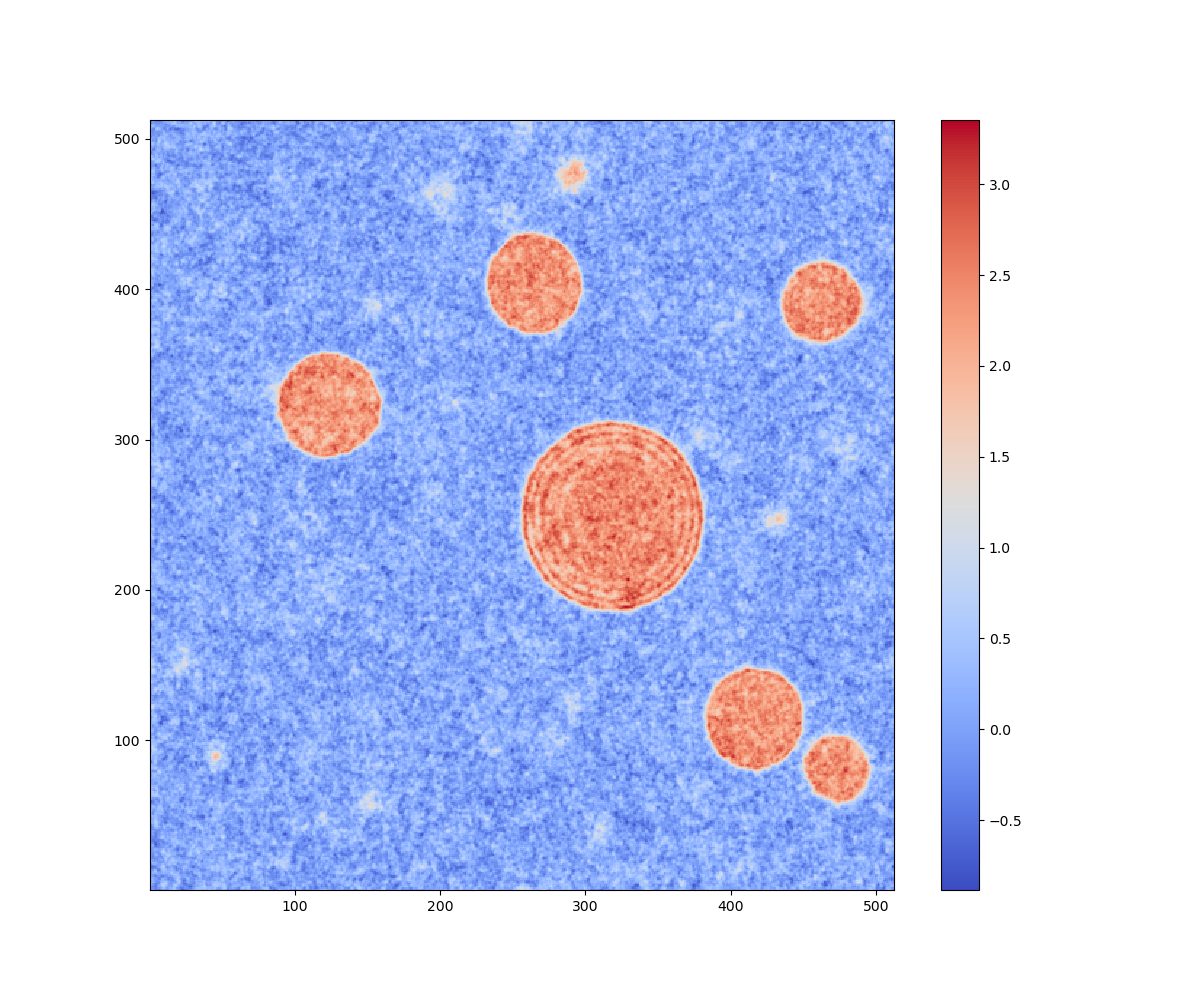}
    \includegraphics[width=0.225\textwidth]{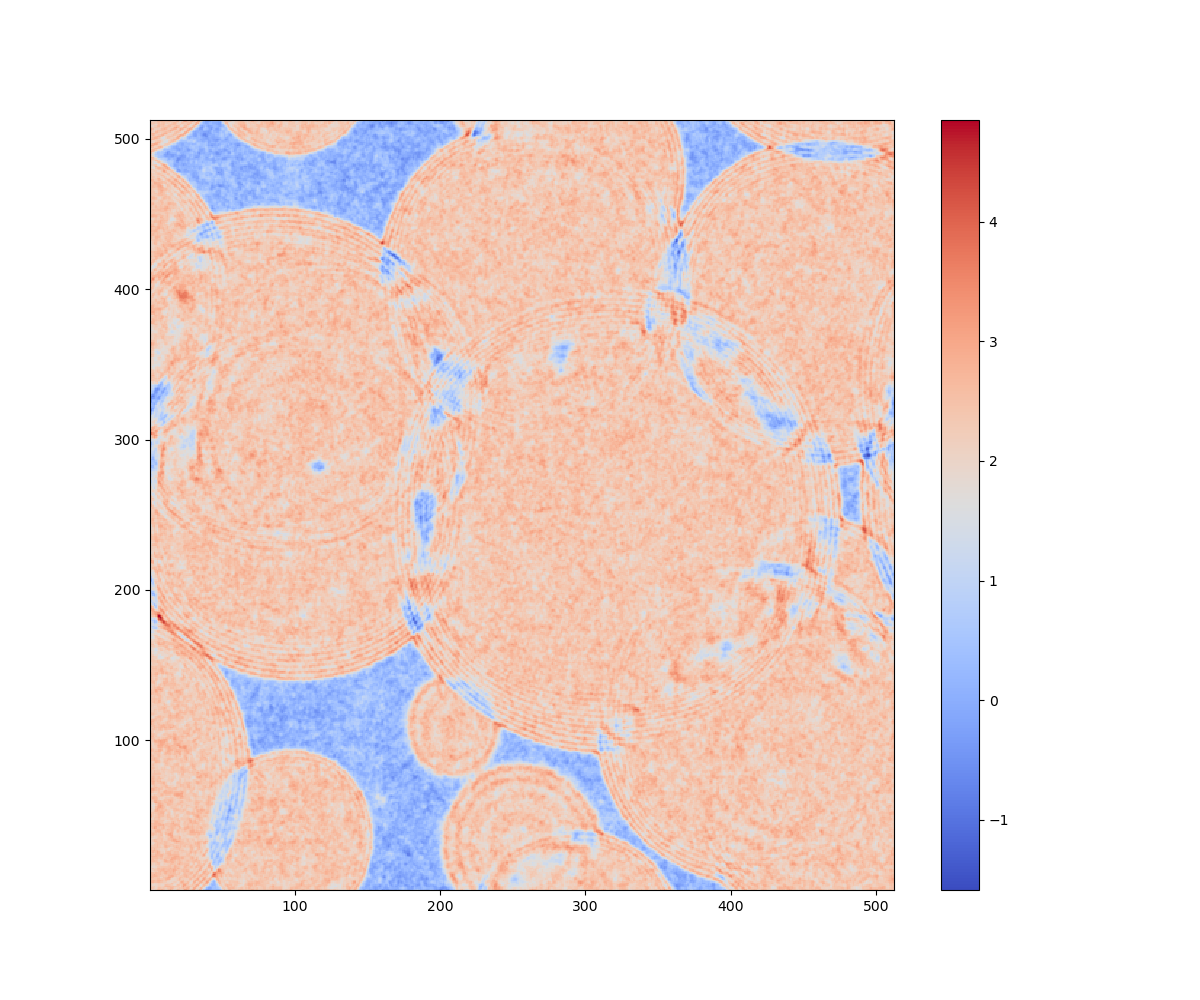}
    \\
    \includegraphics[width=0.225\textwidth]{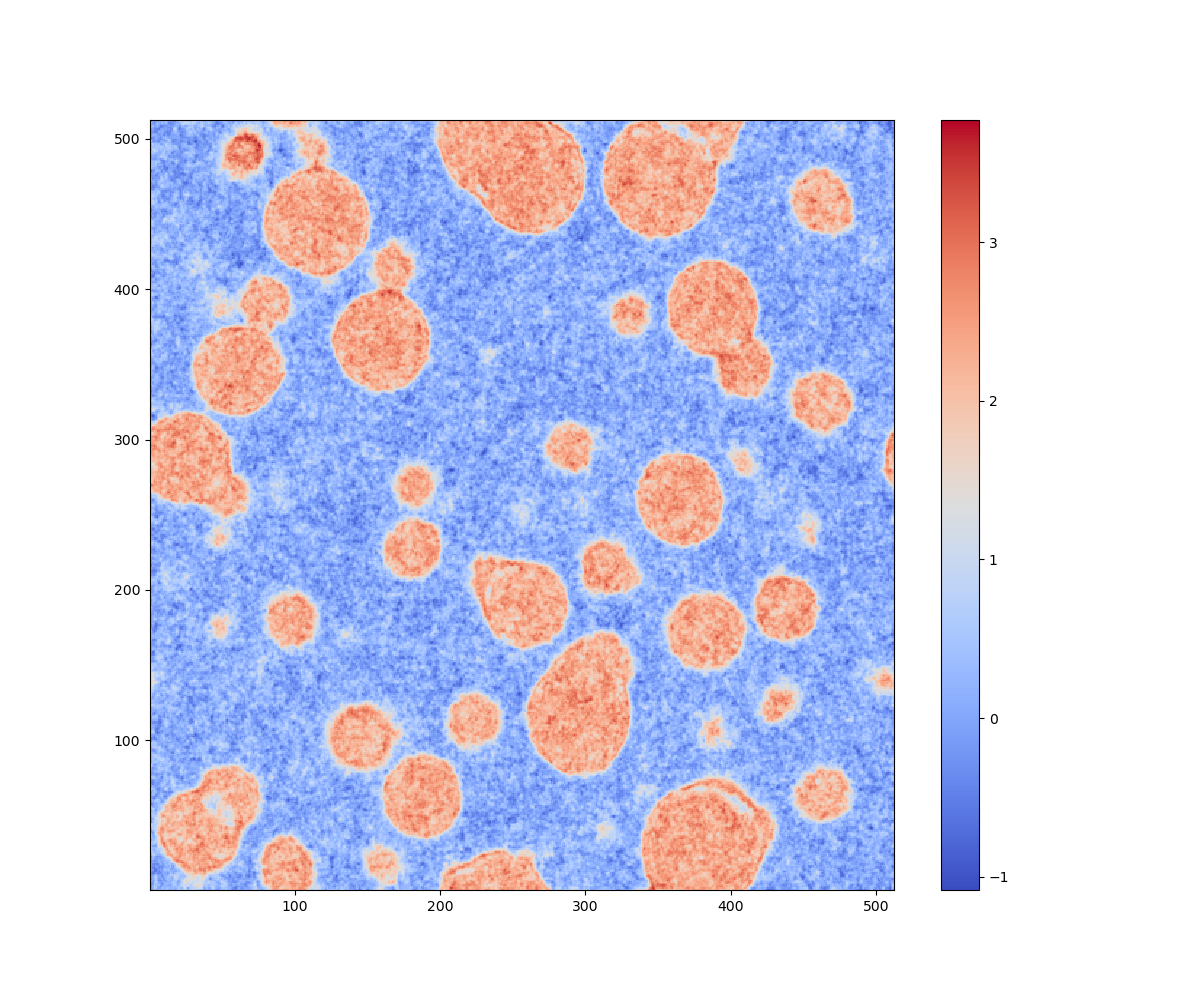}
    \includegraphics[width=0.225\textwidth]{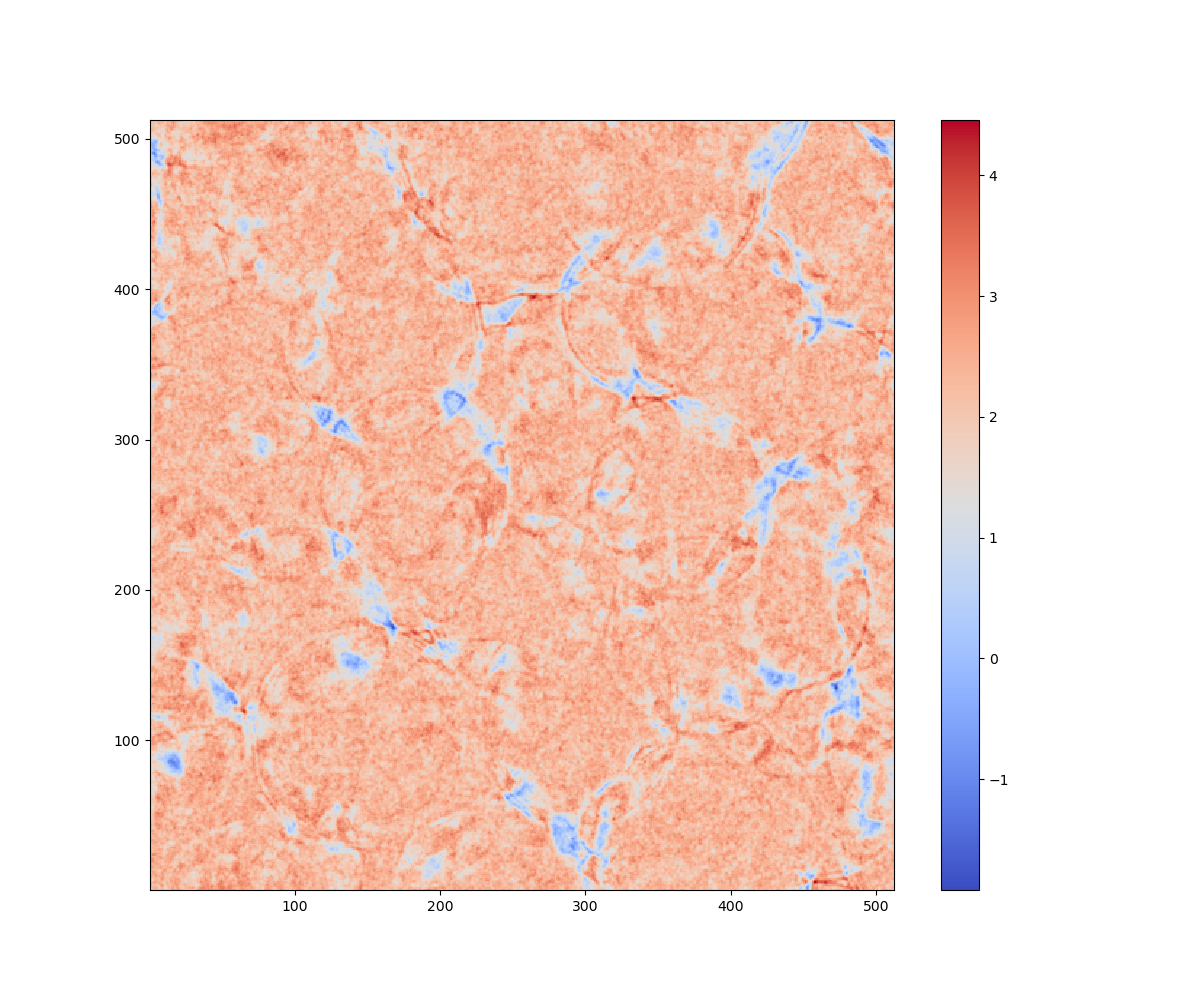}
    \caption{The slices along the $x$ direction of $\phi$ at $\tilde{t}=105.6$(left), $140.8$(right) with the initial temperature $T_i = 1.2M$(top) and at $\tilde{t}=35.2$(left), $52.8$(right) with the initial temperature $T_i = 1.35M$(bottom). Bubbles are formed from thermal fluctuations and then expand and collide with each other.}
    \label{fig:withoutbubble_expand}
\end{figure}

\noindent{\it \bfseries  Numerical results.} 
The initial thermal fluctuations induce a subset of spatial points, then oscillons formed, that first transition to the true vacuum, then triggering adjacent regions to undergo analogous vacuum transitions. These newly transitioned regions subsequently propagate the phase transition to their own neighborhoods, ultimately forming an expanding bubble structure. Afterwards, these bubbles expand and collide with each other and ultimately complete the phase transition.
As an illustration, the slice of the scalar field $\phi$ at different time is shown in Fig.\ref{fig:withoutbubble_expand}, where one can find that a larger number of oscillons and bubbles is formed for a higher initial temperature. 
See Fig.~\ref{fig:3d_snapshot} in the {\it Supplemental Material} for the 3D snapshot of $\phi$.  

\begin{figure}[!htp]
    \centering
    \includegraphics[width=0.4\textwidth]{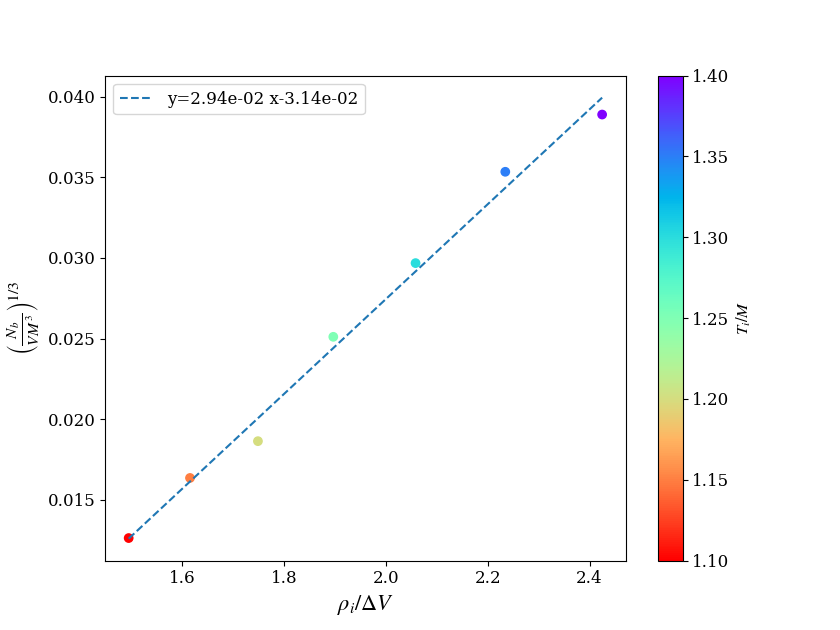}
    \caption{ Bubble number density versus initial energy density ratio ($\Delta V$ is the potential barrier) under different $T_i$, and the dashed line is the linear fitting result.}
    \label{fig:initV_Nb}
\end{figure}

In numerical simulations, the algorithm can count a bubble after it being successfully nucleated through quantum tunneling. However, in the case of thermal fluctuations, the nucleation of bubbles is spontaneous and continuous, so we need an algorithm to identify and count bubbles.
We arrange a series of primary detection points uniformly spaced on the lattice, with one secondary detection point positioned between each pair of adjacent primary points. A potential bubble nucleation is identified when the $\phi$-field value at a primary detection point exceeds the critical value. To eliminate counting errors caused by bubble expansion, if any of the six surrounding primary detection points already contains a bubble and the $\phi$-value at the intervening secondary detection point also surpasses the critical value, the observed $\phi$-value elevation at the current primary point is attributed to the expansion of existing bubbles rather than the formation of a new nucleation site(see details in {\it Supplemental Material}).
The relationship between the number of bubbles $N_b$ obtained by the above algorithm and the initial energy density $\rho_{i} = \frac{1}{2}\dot{\phi}^2+\frac{1}{2}(\nabla\phi)^2+V(\phi)$ at different temperatures is shown in Fig.~\ref{fig:initV_Nb}.
The bubble number density appears to depend linearly on the energy ratio between the initial energy density and the potential barrier. 

\begin{figure}[!htp]
    \centering
    \includegraphics[width=0.4\textwidth]{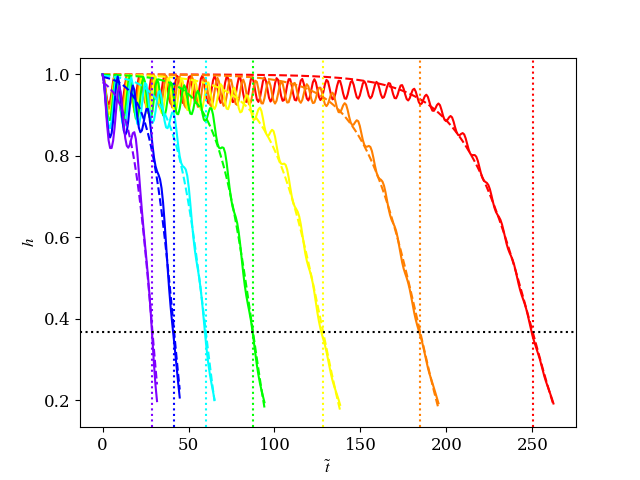}
    \caption{The false vacuum fraction (solid line) and the fitting result (dashed line) in simulation. The vertical dotted line corresponds to the time $t=t_f$ and the horizontal dotted line corresponds to $h=1/e$. The initial temperatures are $T_i=(1.4,1.35,1.3,1.25,1.2,1.15,1.1)$M from left to right curves. } 
    \label{fig:PFVl}
\end{figure}
The system transits to the true vacuum after a period of oscillation in the false vacuum, whose timescale depends on the initial temperature $T_i$. As shown in Fig. \ref{fig:PFVl}, A higher initial temperature leads to a shorter oscillation period of the scalar field.
Like the quantum tunneling case~\cite{Ajmi:2022nmq}, we use the typical function $h_{FV}^{th}=\exp(-\exp(\beta(t-t_f)))$ (the dashed line in Fig. \ref{fig:PFVl}) to fit the false vacuum fraction of the thermal fluctuation one, where $t_f$ marks the time when the phase transition ends, at which point $h=1/e$. In the simulation by assuming the field space is divided into two parts $\phi=0$ and $\phi = \phi_v$, the fraction can be repeated as $h_{FV}^{sim}=1-{\langle\phi\rangle}/{\phi_v}$ (the solid line in Fig. \ref{fig:PFVl}). 

\begin{figure}[!htp]
    \centering
    \includegraphics[width=0.4\textwidth]{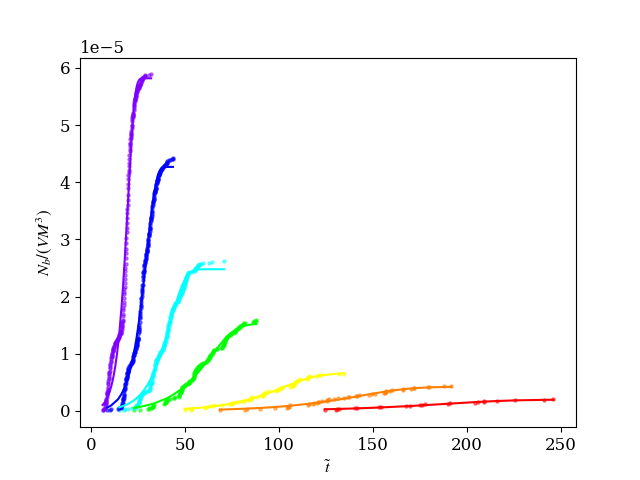}
    \caption{The temporal evolution of bubble number density and its theoretical fit (solid line).}
    \label{fig:nb1}
\end{figure}

With the help of false vacuum fraction $h(t)$, we present and fit the bubble number density $n_b(t)$ evolving over time in Fig.~\ref{fig:nb1}. The fitting line is based on $n_b(t)=\int p(t')h(t')\mathrm{d}t'$ with the nucleation rate expressed as $p(t) = p_f \exp[{\beta (t- t_f)}]$.

\begin{figure}[!htp]
    \centering
    \includegraphics[width=0.4\textwidth]{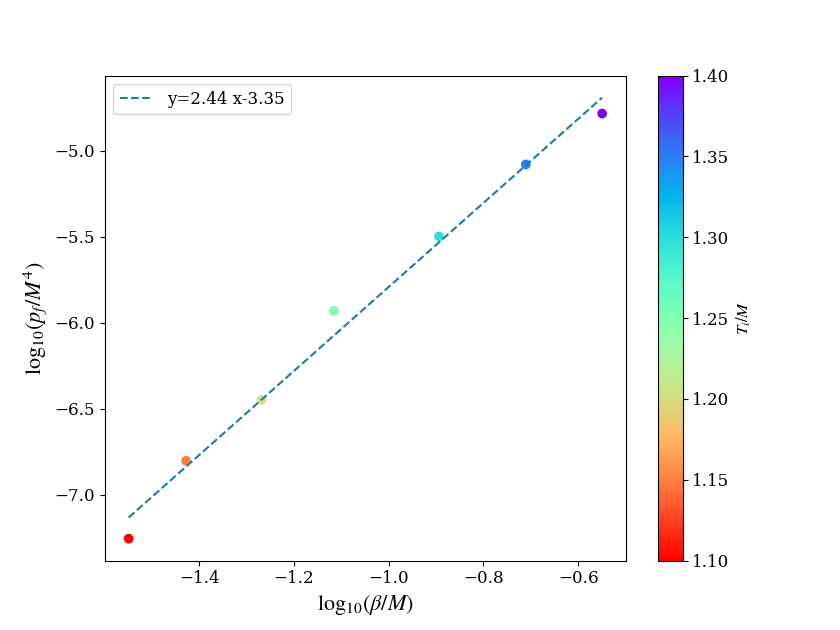}
    \caption{The relationship between $p_f$ and $\beta$ with different $T_i$. The dashed line corresponds to the theoretical fit.}
    \label{fig:pf_beta}
\end{figure}
\begin{figure}[!htp]
    \centering
    \includegraphics[width=0.4\textwidth]{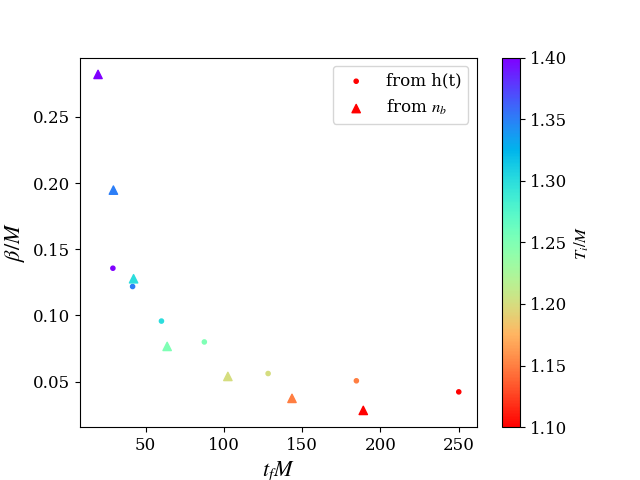}
    \caption{The value of $\beta$ and $t_f$ extracted from $h(t)$ and $n_b(t)$.}
    \label{fig:nb2}
\end{figure}

From the fitting in Fig.~\ref{fig:nb1}, we found that $p_f$ changes with temperature. Therefore, we present the data in Fig. \ref{fig:pf_beta} and fit it using the power law relation $p_f/M^4 \approx (\beta/M)^{2.44}$. It can be seen that both $p_f$ and $\beta$ exhibit growth with increasing $T_i$. The expressions for $h(t)$ and $n_b(t)$ both have parameters $\beta$ and $p_f$, so we can extract these parameters from two different fits and compare them. The result is shown in Fig.~\ref{fig:nb2}.
We found that the discrepancy between the fitted results from $h(t)$ and from the bubble nucleation theory increases as the initial temperature increases, which is partially sourced by the bubble collision dynamics and the thermal fluctuation effects.

\begin{figure}[!htp]
    \centering
    \includegraphics[width=0.4\textwidth]{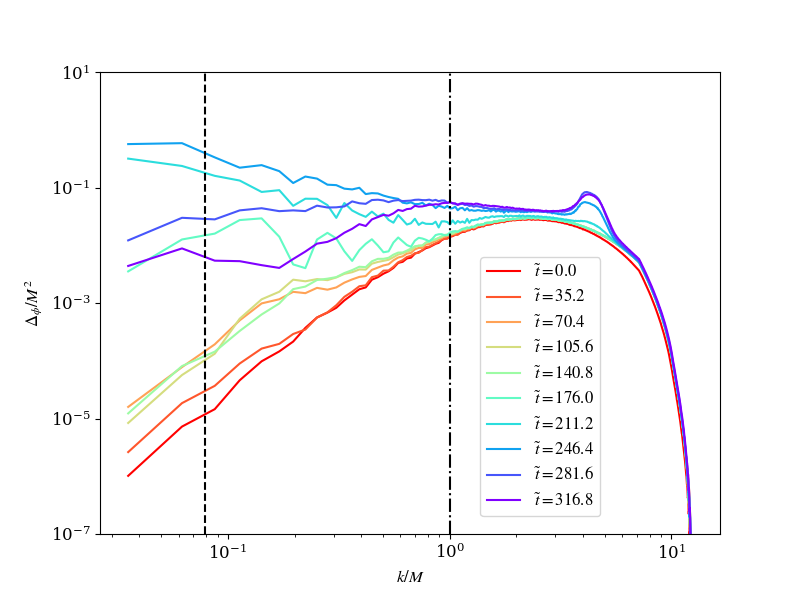}
    \\
    \includegraphics[width=0.4\textwidth]{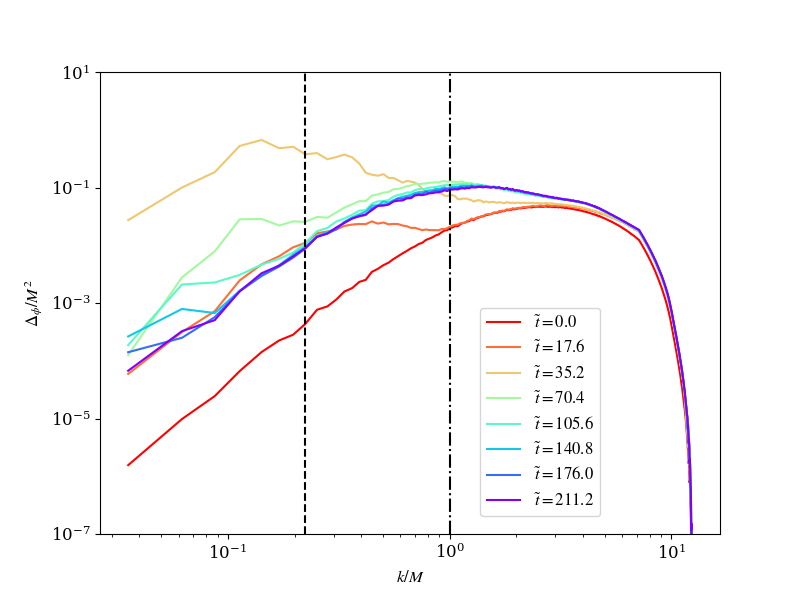}
    \caption{The power spectrum of the scalar field under the initial temperature $T_i = 1.1M(\text{top}), 1.35M(\text{bottom})$. The vertical dashed and dash-dotted lines correspond to $k/M=2 \pi 
 R_*^{-1}$ and $k=M$.}
    \label{fig:scalar_spec}
\end{figure}

The spectra of $\phi$ under different initial temperature are shown in Fig.\ref{fig:scalar_spec}. In the early stages of simulation, oscillons and bubbles begin to form, and the infrared end of the power spectrum slowly rises. More oscillons are formed for higher initial temperature, and yield more significant uplift. When the bubbles expand and begin to collide with each other, the infrared end rises more rapidly. After the phase transition is completed, the spherical symmetry of the bubble is disrupted, and the scalar field enters the true vacuum oscillation stage, with a slight drop in the infrared end. The two typical length, scale of the mean bubble separation at the collision stage $R = (V/N_b)^{1/3}$ and $M^{-1}$, are presented as well. As in Refs.~\cite{Cutting:2018tjt,Cutting:2020nla}, we also observe a small peak in the UV end during the the oscillation stage after bubble collisions, and the peak disappears when the thermal effect increases at a higher initial temperature $T_i$. 





\begin{figure}[!htp]
    \centering
    \includegraphics[width=0.4\textwidth]{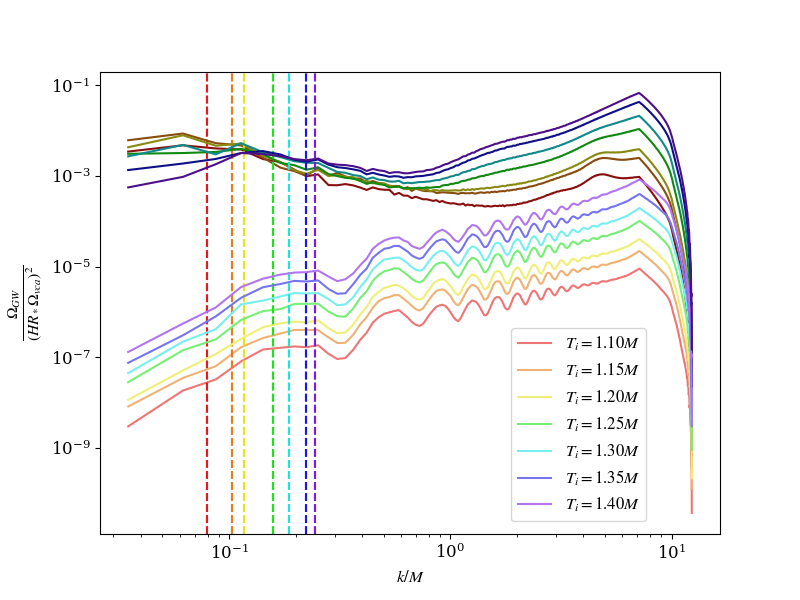}
    \caption{The power spectrum of GWs in early time $\tilde{t} = 8.8$ (light color) and the end of the simulation (deep color), where the vertical dashed lines correspond to the mean bubble separation $2\pi R_*$. }
    \label{fig:gws}
\end{figure}

Finally, we found that the PT dynamics under study can indeed induce GW productions. The GW spectra at different temperatures is given by Fig. \ref{fig:gws}, where the amplitudes in IR part arising from the bubble collisions grow by around 3-4 orders, and the peak in the high wave number regions comes from the characteristic bubble wall thickness. The magnitude of the GW spectra decreases by smaller amount for higher initial temperature, because more bubbles are nucleated. The results show that the amplitude of the GW spectra $\Omega_{GW}\approx \mathcal{O}(10^{-3})(H R_*\Omega_{vac})^2$, where $\Omega_{vac} = \Delta V / \rho_r$ is the ratio of $\Delta V$ to the radiation energy density $\rho_r$. This behavior aligns with the quantum tunneling scenario of the vacuum PT studied in Refs.~\cite{Cutting:2018tjt,Cutting:2020nla}.

\noindent{\it \bfseries  Conclusion and discussion.}
In this Letter, we have found that the first-order PT can occur through the thermal fluctuation, and observed that the probability of formation of oscillons and further nucleated vacuum bubbles increases as the initial energy density grows higher than the potential barrier. Our analysis have demonstrated that the PT proceeds through four distinct dynamical stages: (1) formation of the oscillons, (2) nucleation of vacuum bubbles, (3) subsequent expansion, and (4) eventual bubble collision. More importantly, we have quantitatively established that the false vacuum fraction follows a well-defined functional which depends on the PT duration parameter $\beta$, and is in precise agreement with theoretical predictions from the first-order PT nucleation theory. We have also found that the first-order PT process driven by the thermal fluctuation can produce a significant amount of GWs with the peak weak number being characterized by the mean bubble separation $R_*$.



\noindent{\it \bfseries Acknowledgements.}
The numerical calculations in this study were carried out on the ORISE Supercomputer.
This work is supported by the National Key Research and Development Program of China under Grant No. 2021YFC2203004.
L.B. is supported by the National Natural Science Foundation of China (NSFC) under Grants Nos.  12322505, and 12347101.
L.B. also acknowledges Chongqing Natural Science Foundation under Grant
No. CSTB2024NSCQ-JQX0022 and 
Chongqing Talents: Exceptional Young Talents Project No. cstc2024ycjh-bgzxm0020.

\bibliography{reference.bib}

\clearpage

\onecolumngrid
\begin{center}
  \textbf{\large \it Supplemental Material}\\[.2cm]
\end{center}

\onecolumngrid
\setcounter{equation}{0}
\setcounter{figure}{0}
\setcounter{table}{0}
\setcounter{section}{0}
\setcounter{page}{1}
\makeatletter
\renewcommand{\theequation}{S\arabic{equation}}
\renewcommand{\thefigure}{S\arabic{figure}}

This supplemental material includes the details of our simulation and explanations of the results presented in the main text, as well as some extra results. We begin with a detailed description of our numerical scheme, including equations of motion and initial conditions. 

\section{Equations of motion}\label{sec:SupEOM}
\begin{figure}[!htp]
    \centering
    \includegraphics[width=0.4\linewidth]{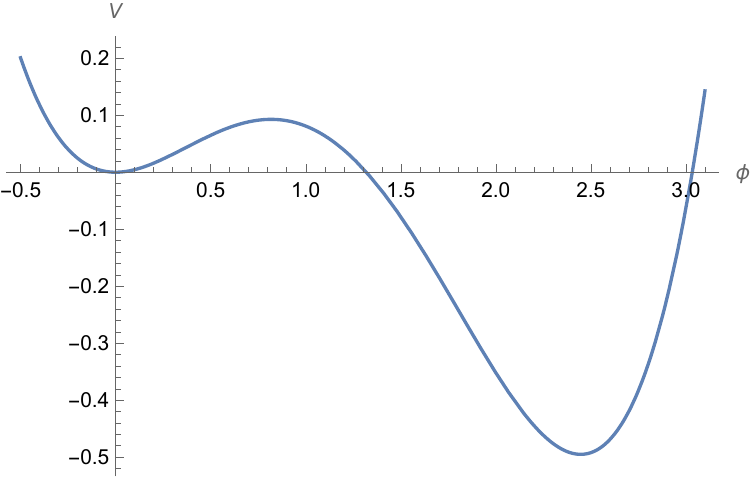}
    \caption{Shape of the scalar potential}
    \label{fig:shapeofpotential}
\end{figure}

We conduct simulations in Minkovsky space-time.
The field equations of motion can be obtained by varying the following action
\begin{equation}
S=-\int d^4x[\frac{1}{2}\partial^\mu\phi\partial_\mu\phi+V(\phi)]
\end{equation}
with $\phi$ the scalar singlet. The potential we considered is 
\begin{eqnarray}
V(\phi)=\frac{1}{2}M^2\phi^2+\frac{1}{3}\delta\phi^3+\frac{1}{4}\lambda\phi^4\;.
\end{eqnarray}
 For the shape of the potential under study, see FIG. \ref{fig:shapeofpotential}.

We use two parameters $f_*$ and $w_*$ to do the rescale for physical quantities
\begin{equation}
\tilde{\phi}=\phi/f_*, \quad {\rm d}\tilde{t}=w_*{\rm d}t, \quad {\rm d}\tilde{x}^i=w_*{\rm d}x^i, \quad \tilde{S}=\left(\frac{w_*}{f_*}\right)^2S(f_*\tilde{\phi}). \label{dimensionless rescale} 
\end{equation}
For convenience, we set $f_*=w_*=150$ GeV. Then, the equations of motion expressed by dimensionless fields and space-time variables follow immediately from varying the dimensionless action $\tilde{S}$ \begin{eqnarray}
\Ddot{\widetilde{\phi}}-\widetilde{\nabla}^2\widetilde{\phi} + \widetilde{V}_{,\widetilde{\phi}} = 0 
\end{eqnarray}
with $\widetilde{V}_{,\widetilde{\phi}}=\partial \widetilde{V}(\widetilde{\phi}) / \partial \widetilde{\phi},$  $\dot{} ={\rm d}/{\rm d}\tilde{t}$ and $\widetilde{\nabla_i}=\partial/\partial \tilde{x}^i$ here.

\section{Initial Conditions} 
\label{sec:initial}
In our simulation, the scalar field can be considered to be in thermal equilibrium. We can use the following thermal spectrum to describe the amplitude and momentum distribution of $\phi$ in momentum space
\begin{equation}
\mathcal{P}_{\phi}(k)=\frac{n_k}{w_k}=\frac{1}{w_k}\frac{1}{e^{w_k/T}-1}, \quad \mathcal{P}_{\dot{\phi}}(k)=n_k w_k=\frac{w_k}{e^{w_k/T}-1},
\end{equation}
where $n_k$ denotes the occupation number of the Bose-Einstein distribution, $w_k=\sqrt{k^2+m_{\rm eff}^2}$ and $k$ are physical frequency and momenta respectively, $m_{\rm eff}^2$ is the initial effective mass square of scalar field and overdots represent differentiation with respect to cosmic time $t$.

For continuum, the two-point correlation functions can be written as
\begin{align}
\langle \phi (\boldsymbol{{\rm k}}) \phi(\boldsymbol{{\rm k}}') \rangle &= (2\pi)^3\mathcal{P}_{\phi}(k)\delta(\boldsymbol{{\rm k}}-\boldsymbol{{\rm k}}'),\\
\langle \dot{\phi} (\boldsymbol{{\rm k}}) \dot{\phi}(\boldsymbol{{\rm k}}') \rangle &= (2\pi)^3\mathcal{P}_{\dot{\phi}}(k)\delta(\boldsymbol{{\rm k}}-\boldsymbol{{\rm k}}'), \nonumber \\
\langle \phi(\boldsymbol{{\rm k}}) \dot{\phi}(\boldsymbol{{\rm k}}') \rangle &= 0 \nonumber \,.
\end{align}
With proper rescale, we can recreate the correlation functions equivalent to that in the continuum on the discrete lattice, which does not depend explicitly on the volume 
\begin{align}
\qquad \langle | \phi(\boldsymbol{{\rm k}}) |^2 \rangle &= \left(\frac{N}{\delta x_{\rm phy}}\right)^3\mathcal{P}_{\phi}(k), \qquad \langle \phi(\boldsymbol{{\rm k}})  \rangle = 0, \\
\qquad \langle | \dot \phi(\boldsymbol{{\rm k}}) |^2 \rangle &= \left(\frac{N}{\delta x_{\rm phy}}\right)^3\mathcal{P}_{\dot{\phi}}(k),\qquad \langle \dot \phi(\boldsymbol{{\rm k}}) \rangle = 0, \nonumber
\end{align}
where $N$ denote the number of points per side and $\delta x_{\rm phy}$ denote the physical lattice spacing. We first generate $\phi(\boldsymbol{{\rm k}})$ and $\dot \phi(\boldsymbol{{\rm k}})$ following Gaussian random distribution in momentum space, which contain all modes from infrared truncation $k_{\rm IR}$ to maximum momentum $k_{\rm max}=\sqrt{3}k_{\rm UV}$ of the simulation box, with $k_{\rm UV}$ the ultraviolet truncation in one direction. Then, by applying the discrete inverse Fourier transform to the field in momentum space, we can acquire the field in three-dimensional coordinate space.

\begin{figure}[!htp]
    \centering
    \includegraphics[width=0.35\linewidth]{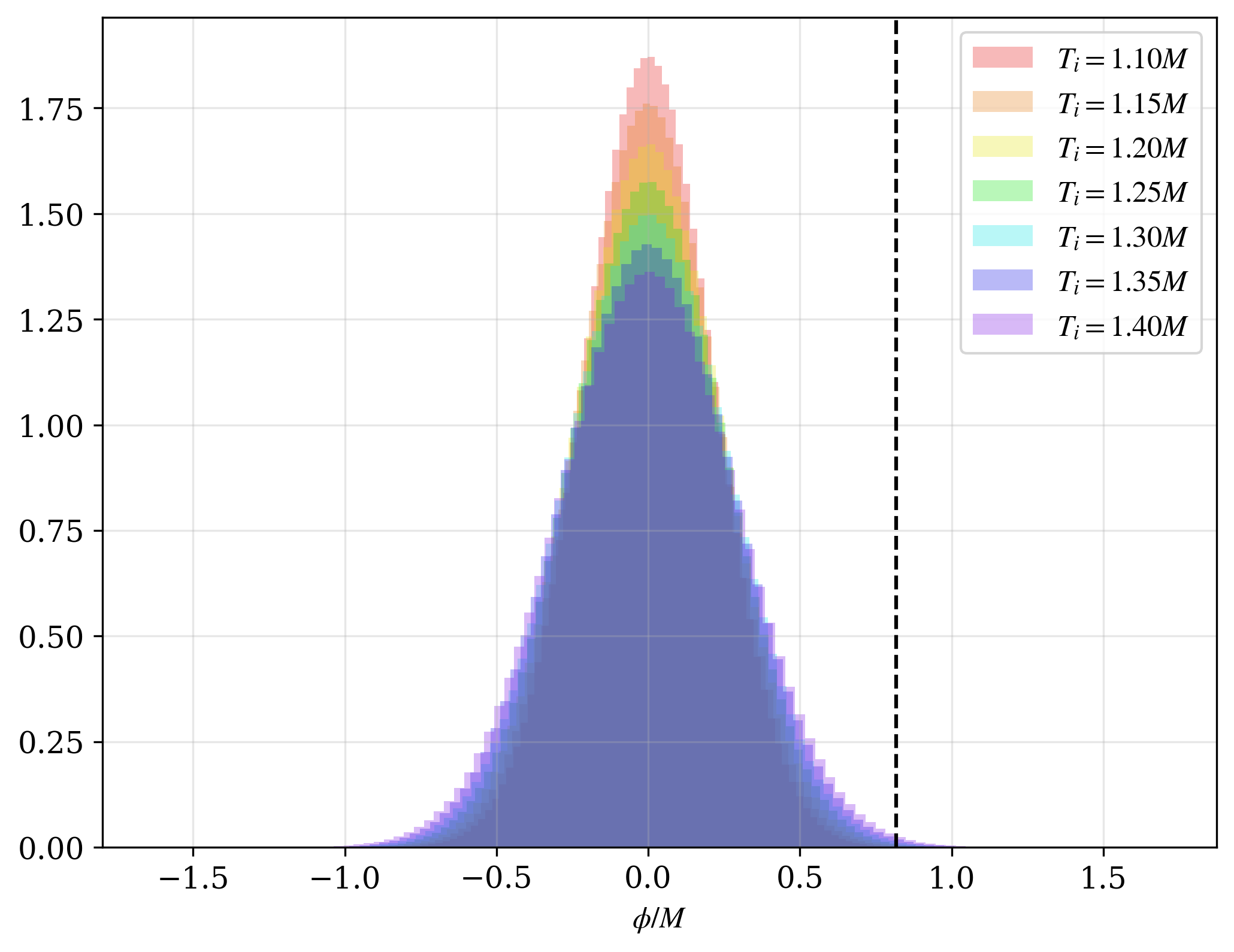}
    \includegraphics[width=0.4\linewidth]{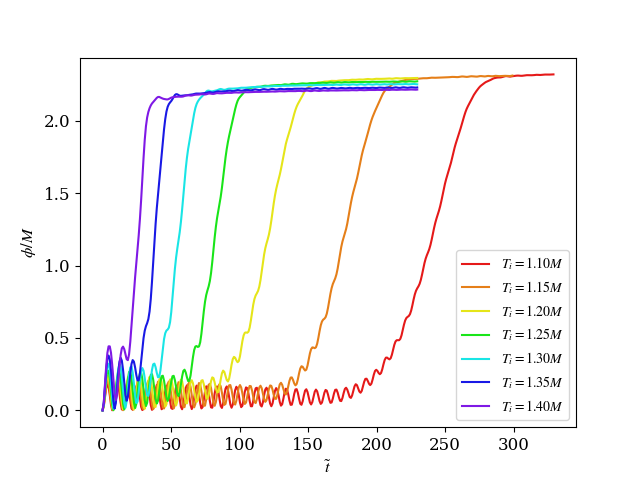}
    \caption{The initial scalar filed distributions(left), where the black dashed line is the field value correspond to potential barrier, and the evolution of the average field value(right) in the lattice simulation.}
    \label{fig:phi_dis}
\end{figure}

We show the initial field distribution and the averaged scalar field value during the PTs in Fig.\ref{fig:phi_dis}. The initial thermal fluctuations induce a subset of spatial points where the $\phi$-field values overcome the potential barrier, resulting in $dV/d\phi < 0$ at these locations. Within the equation of motion governing the $\phi$-field dynamics, this condition leads to $\dot{\pi} > 0$, thereby accelerating the $\phi$-field's evolution toward the true vacuum at these locations. When the $\phi$-field value at a spatial point approaches sufficiently close to the true vacuum, neighboring regions where the $\phi$-field remains on the left of the potential barrier (characterized by $dV/d\phi > 0$) develop substantial field gradients. These gradients satisfy the inequality $\nabla^2\phi - dV/d\phi > 0$, which similarly induces $\dot{\pi} > 0$. 
We can see that the field arises fast with the temperature increase.

\section{ Gravitational wave spectrum}
The GWs is the metric perturbation around flat-space metric with the definition of $h_{\mu \nu} = g_{\mu \nu} - \eta_{\mu \nu}$. Expanding the Einstein equation to the linear order in $h_{ij}$ and imposing the transverse-traceless(TT) gauge, the GWs field equation of motion is derived as
\begin{equation}
    \ddot{h}_{ij} - \nabla^2 h_{ij} = \frac{2}{m_{pl}^2}\Pi^{TT}_{ij}	
    \label{eq:gws} \;,
\end{equation}
where $\dot{} = \partial / \partial t$, $ \boldsymbol{\nabla} = \partial / \partial x_i $, $m_{pl}$ is the reduced Planck energy and $\Pi^{TT}_{ij}$ is the transverse and traceless part of anisotropic tensor $\Pi_{ij}$. The anisotropic stress tensor is given by the energy momentum tensor $T_{ij}$ with respect to a perfect fluid $\Pi_{ij} = T_{ij} - pg_{ij}$, where $T_{ij} = \partial_i \phi \partial_j \phi$ in this case and the second term background pressure $ p = \sum_{i} T_{ii}/3$ does not contribute to $\Pi^{TT}_{ij}$. 
The GWs source $\Pi^{TT}_{ij}$ is extracted in Fourier space by $\Pi_{ij}^{TT} = \Lambda_{ij,kl} \Pi_{kl}$, where the projection operator $\Lambda_{ij,kl}$ is defined as
\begin{align}
    \Lambda_{ij,kl}(\hat{\textbf{k}}) &= P_{ik}({\hat{\textbf{k}}}) P_{jl}({\hat{\textbf{k}}})-\frac{1}{2}P_{ij}({\hat{\textbf{k}}})P_{kl}({\hat{\textbf{k}}}) \;,\nonumber \\
    P_{ij}({\hat{\textbf{k}}}) &= \delta_{ij} - \hat{k}_i\hat{k}_j \;,
\end{align}
where $\hat{k}_i = k_i / k$ is the unit vector in $\textbf{k}$ direction.
For convenience in the numerical calculation of gravitational waves (GWs), an auxiliary tensor $u_{ij}$ is introduced and dynamically evolved, where the metric perturbation is derived via the projection $h_{ij} = \Lambda_{ij,kl} u_{ij}$. The equation of motion governing $u_{ij}$ takes the form  
\begin{align}
    \ddot{u}_{ij} - \nabla^2 u_{ij} = \frac{2}{m_{pl}^2}\Pi_{ij} \;.
\end{align}
The gravitational wave energy density is calculated by
\begin{align}
    \rho_{gw}(t) = \frac{1}{32 \pi G}\left<\hat{h}_{ij}(\textbf{x},t) \hat{h}_{ij}(\textbf{x},t) \right> \;, 
\end{align}
where $\left<...\right>$ is the volume average. In the momentum space, the power spectrum of metric perturbation is expressed by 
\begin{align}
    \frac{\mathrm{d} \rho_{gw}}{\mathrm{d} \log{k}} = \frac{k^3}{(4\pi)^3G} P_{\hat{h}}(k,t) \;,
\end{align} 
where $\left<\hat{h}_{ij}(\textbf{k},t) \hat{h}_{ij}(\textbf{k},t) \right> = (2\pi)^3P_{\hat{h}}(k,t) \delta^{(3)}(\textbf{k} - \textbf{k}^\prime)$, and the normalized GWs power spectrum is  
\begin{align}
\Omega_{GW} = \frac{1}{\rho_c}\frac{\mathrm{d} \rho_{gw}}{\mathrm{d} \log{k}} \;,
\end{align}
where the critical density $\rho_c = \frac{3H^2}{8\pi G}$.

\section{ Bubble Identification Algorithm}\label{sec:bubble_idf}
When investigating First-Order Phase Transitions, the topic of vacuum bubbles is invariably encountered. In lattice simulations, one can induce phase transitions by either placing bubbles at fixed locations or generating them randomly. This approach allows computer programs to easily determine when and where bubbles have formed, as well as their quantity. However, when bubble nucleation occurs spontaneously, such as through thermal fluctuations of a scalar field overcoming a potential barrier, the statistical quantification of bubble formation becomes a non-trivial issue. 

To address this issue, I have developed a program capable of identifying and counting the nucleation of bubbles. This program requires only three additional parameters, imposes minimal computational overhead, and demonstrates robust performance in accurately quantifying randomly generated bubbles.

The methodology begins with the introduction of three key parameters:  
\begin{itemize}
    \item $R$, represents the interval between detection points. It is chosen to be comparable to the radius of a bubble and is divisible by the lattice edge length $N$. 
    \item $\phi_1\sim0.9v$, serves as a threshold: if the modulus of the scalar field $|\phi|$ at a given point exceeds this value, it is identified as a bubble nucleation site.
    \item $\phi_2\sim0.1v<\phi_1$, is introduced to account for the influence of bubble expansion and aids in the discrimination process.
\end{itemize} 

The first step involves collecting data from the detection points to identify potential bubble nucleation sites at the current time step. Here, the detection points are categorized into \textit{primary detection points} ($n_{x,y,z}\mod R=0$) and \textit{secondary detection points} ($n_{x,y,z}\mod (R/2)=0$). The program systematically scans these detection points. When the scalar field value $\phi$ at a \textit{primary detection point} exceeds $\phi_1$, the index $n = n_xN^2 + n_yN + n_z$ is recorded in the array \texttt{BubblePoint}. Similarly, if $\phi$ at a \textit{secondary detection point} exceeds $\phi_2$ , the index $n = -(n_xN^2 + n_yN + n_z)$ is also stored in \texttt{BubblePoint}. For detection points that do not meet these criteria, the value $-1$ is assigned to \texttt{BubblePoint}. This step can be parallelized across multiple cores, where each core computes its local \texttt{BubblePoint} array independently. The results from all cores are then aggregated to the root process using the \texttt{MPI\_Gather()} function, forming a consolidated array \texttt{root\_BubblePoint}. This parallelization strategy significantly enhances computational efficiency. Subsequent calculations are exclusively performed by the root process.

It is important to note that not all coordinate points in \texttt{root\_BubblePoint} correspond to actual bubble nucleation sites. Some points may have been "occupied" due to the expansion of neighboring bubbles. This is the core issue that the subsequent steps of the program aim to resolve. From the previous step, we know that the indices $n$ of all \textit{primary detection points} with $\phi>\phi_1$  are greater than 0. Leveraging this condition, the program systematically examines each \textit{primary detection point} to determine whether it genuinely represents the nucleation of a new bubble.

First, we define a global array \texttt{TrueVacuum}. For each $n>0$, the program checks whether $n$ already exists in \texttt{TrueVacuum}. If it does, the subsequent steps are skipped, and the program proceeds to the next $n$. If $n$ is not present in \texttt{TrueVacuum}, the index $n$ is added to \texttt{TrueVacuum}. This array plays a crucial role in reducing computational overhead and ensuring that no duplicate counting occurs, thereby enhancing the efficiency and accuracy of the algorithm.

Next, the program determines whether the coordinate point $\boldsymbol{n}=(n_x,n_y,n_z)$ that has transitioned to the true vacuum represents the nucleation of a new bubble or has been "occupied" by the expansion of a neighboring bubble. To achieve this, the algorithm examines the 18 \textit{primary detection points} surrounding the coordinate $\boldsymbol{n}$, denoted as $\boldsymbol{n}_\mathrm{neighbour} = (n_x\pm R,n_y,n_z), (n_x,n_y\pm R,n_z), (n_x,n_y,n_z\pm R), (n_x\pm R,n_y\pm R,n_z), (n_x\pm R,n_y,n_z\pm R), (n_x,n_y\pm R,n_z\pm R)$.
For each $\boldsymbol{n}_\mathrm{neighbour}$ , the program first checks whether it exists in the \texttt{TrueVacuum} array. If it does not, the subsequent steps are skipped, and the algorithm proceeds to the next $\boldsymbol{n}_\mathrm{neighbour}$ . If $\boldsymbol{n}_\mathrm{neighbour}$ is present in \texttt{TrueVacuum}, it indicates that this point has already transitioned to the true vacuum. At this stage, the \textit{secondary detection points} play a critical role. These \textit{secondary detection points} are located between two \textit{primary detection point}, and their information is stored in \texttt{root\_BubblePoint} as negative values (excluding $-1$).
The program then searches for the indices of the \textit{secondary detection points} between $n$ and $\boldsymbol{n}$  in \texttt{root\_BubblePoint}. If no such indices are found, it is concluded that $n$ and $\boldsymbol{n}$ reside within two distinct bubbles. Conversely, if \textit{secondary detection points} are identified, it is inferred that both points belong to the same bubble, and the transition at $n$ is attributed to the expansion of the bubble originating from $\boldsymbol{n}_\mathrm{neighbour}$ . We classify this scenario as Case Expansion.

The above steps are repeated until all 18 $\boldsymbol{n}_\mathrm{neighbour}$  points have been evaluated. A new bubble is confirmed to have nucleated at $\boldsymbol{n}$ only if none of the 18 $\boldsymbol{n}_\mathrm{neighbour}$  satisfy Case Expansion (i.e., $\boldsymbol{n}_\mathrm{neighbour}$ and the \textit{secondary detection points} between $\boldsymbol{n}$ and $\boldsymbol{n}_\mathrm{neighbour}$ both transitioned to the true vacuum). This stringent criterion ensures the accurate identification of new bubble nucleation sites while effectively excluding cases where the transition is caused by the expansion of neighboring bubbles.

Finally, it is essential to note that in lattice simulations, periodic boundary conditions are commonly employed. When searching for neighboring \textit{primary detection points}, situations may arise where the search extends beyond the boundaries of the lattice. For instance, if $n_x+R>N$, the actual point identified is $n_x+R-N$. Similarly, if $n_x-R<0$, the point identified is $n_x-R+N$. Therefore, a more general approach is adopted: for a given index $n$, the program first reconstructs the coordinates $n_x = n\mod N^2, n_y = (n\mod N^2)/N, n\mod N$ from $n$ (Here, the symbol $/$ denotes integer division.). Then, to ensure that $\boldsymbol{n}_\mathrm{neighbour}$  remains within the valid range of the lattice, the coordinates are adjusted using $n_{x,y,z\mathrm{neighbour}} = (n_{x,y,z}\pm R+N)\mod N$. Finally, the adjusted coordinates are converted back into the index $n_\mathrm{neighbour}$ , which is then used to search within the \texttt{TrueVacuum} array. 
This periodic wrapping ensures the continuity of the lattice and is a critical consideration in the accurate implementation of the algorithm.

\begin{figure*}[!htp]
    \centering
    \includegraphics[width=0.3\linewidth,trim = 200 70 135 105, clip]{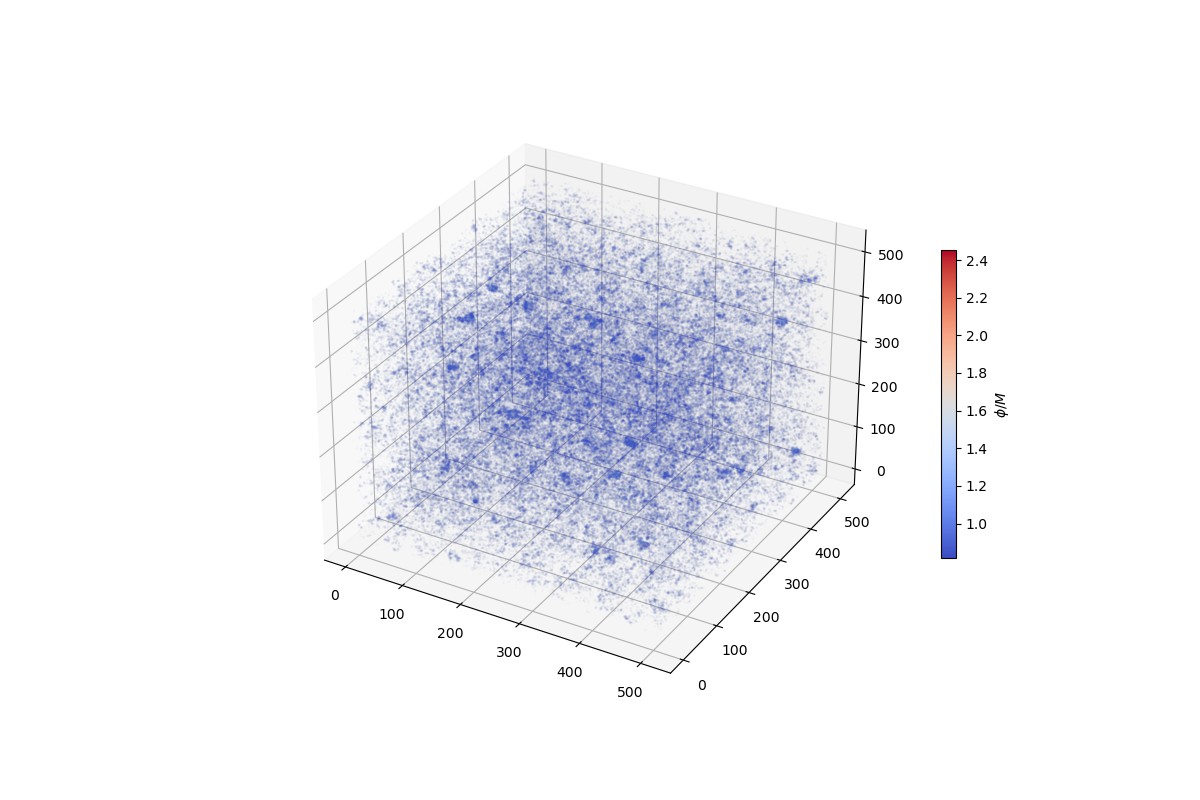}
    \includegraphics[width=0.3\linewidth,trim = 200 70 135 105, clip]{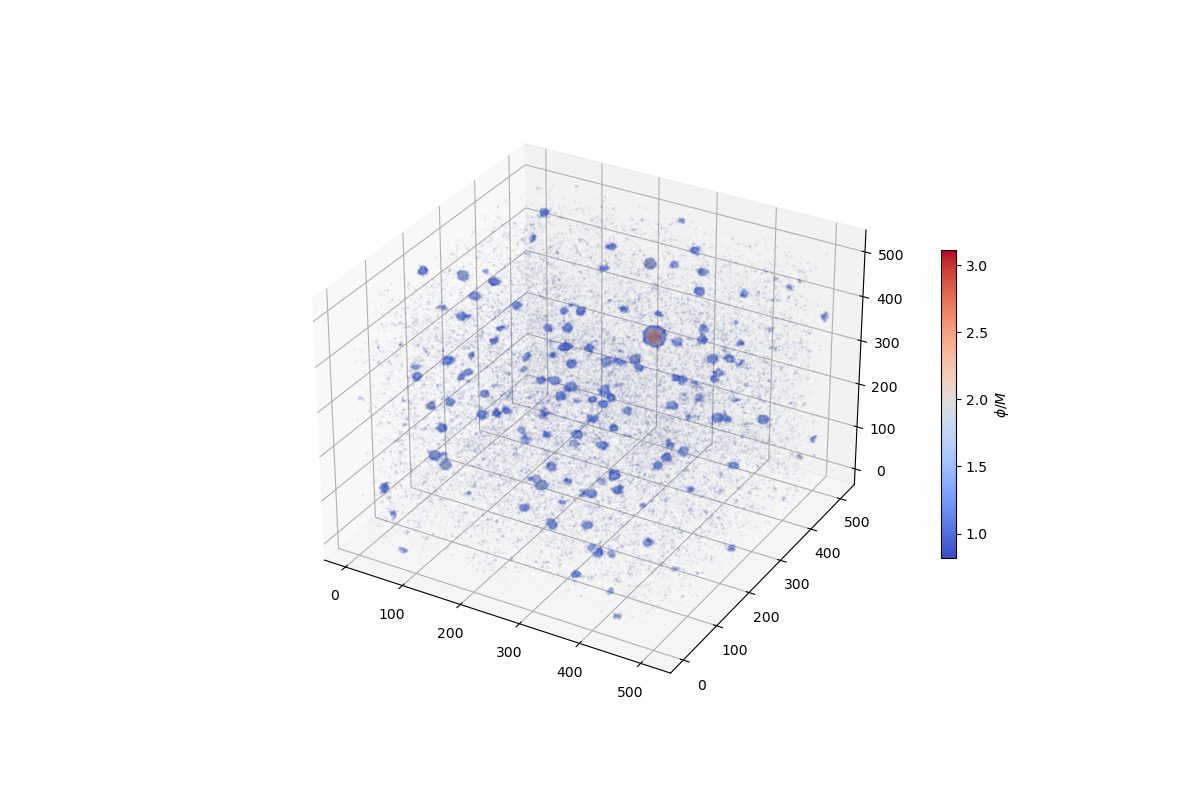}
    \includegraphics[width=0.3\linewidth,trim = 200 70 135 105, clip]{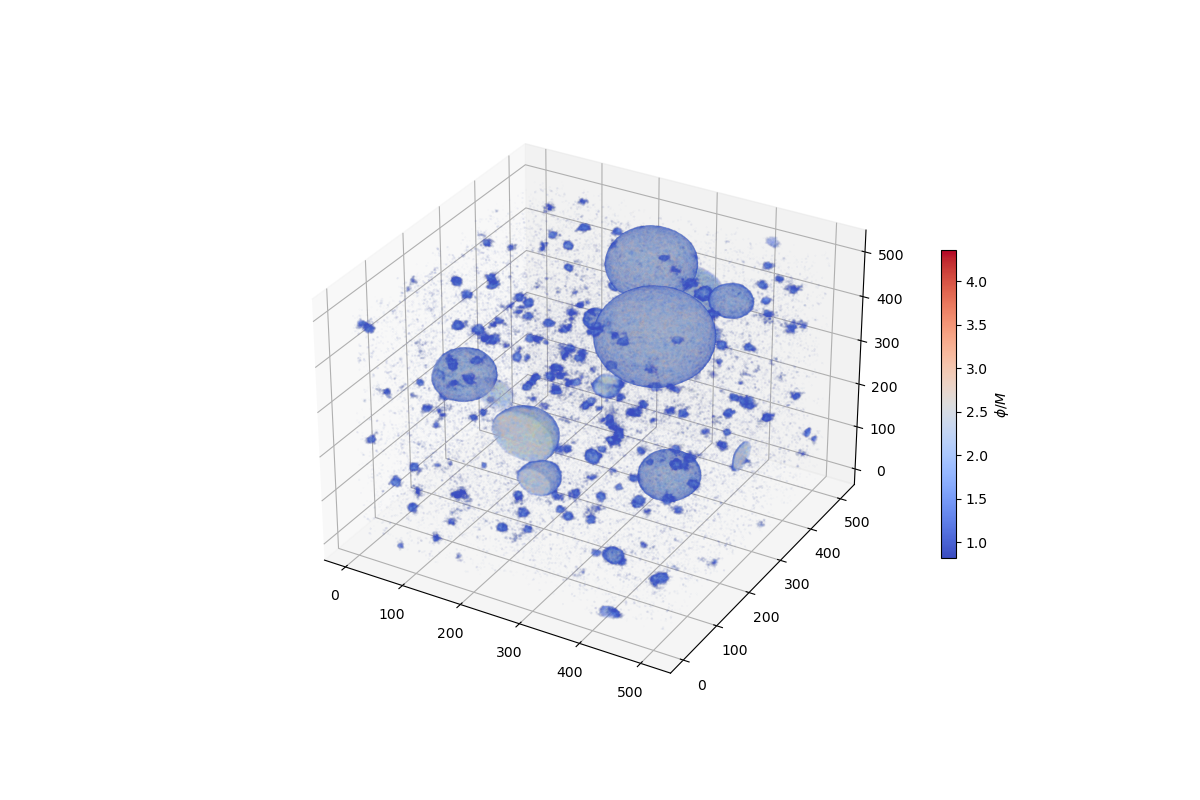}
    \\
    \includegraphics[width=0.3\linewidth,trim = 200 70 135 105, clip]{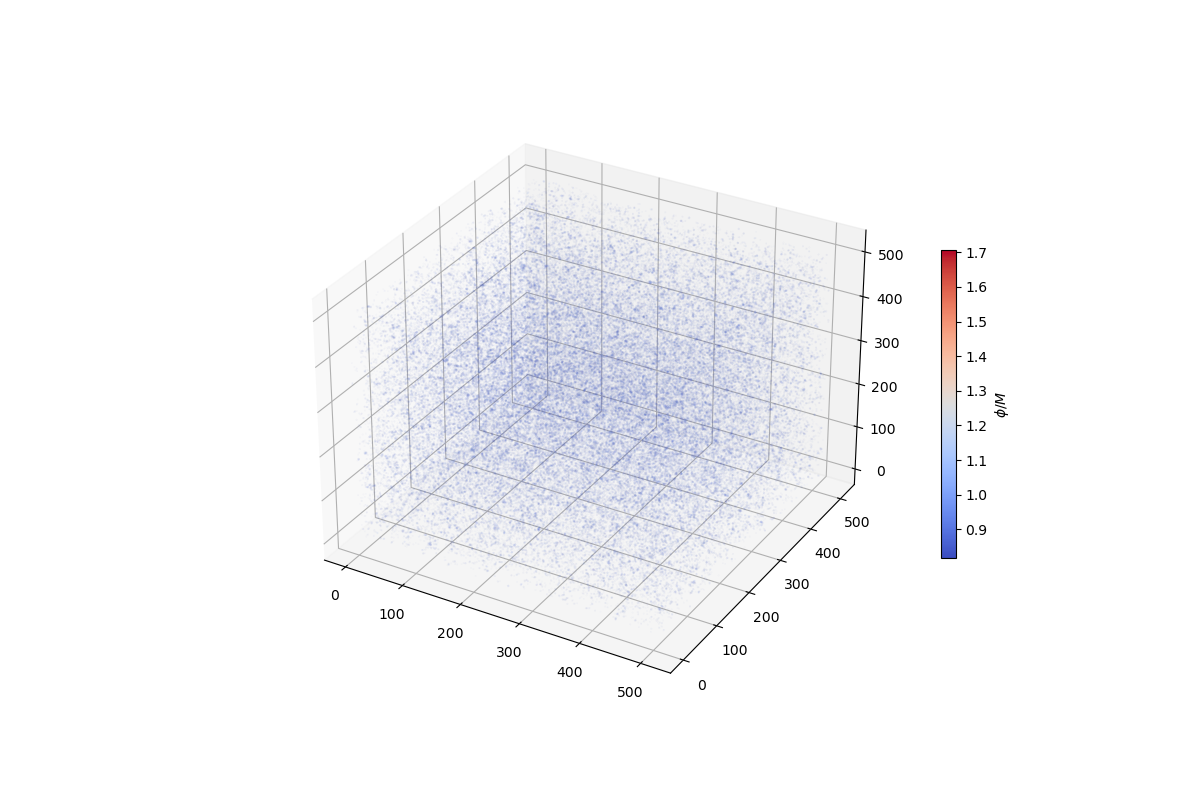}
    \includegraphics[width=0.3\linewidth,trim = 200 70 135 105, clip]{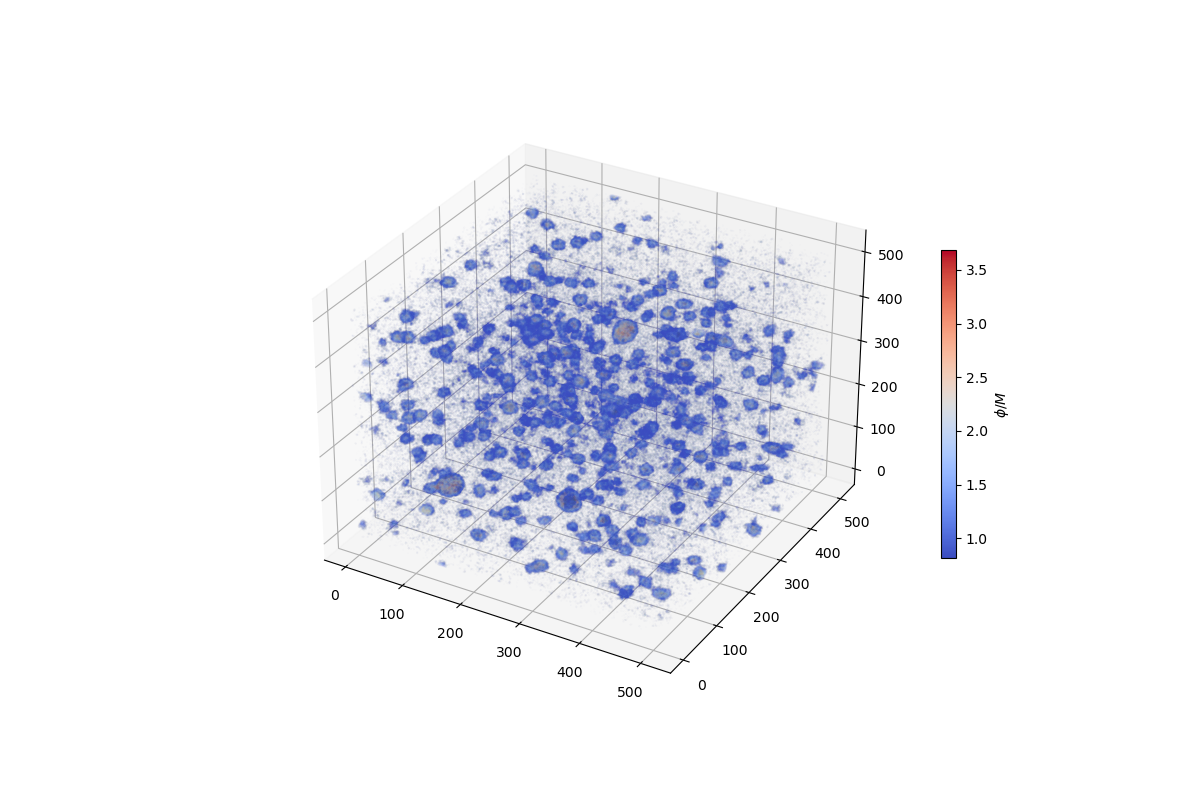}
    \includegraphics[width=0.3\linewidth,trim = 200 70 135 105, clip]{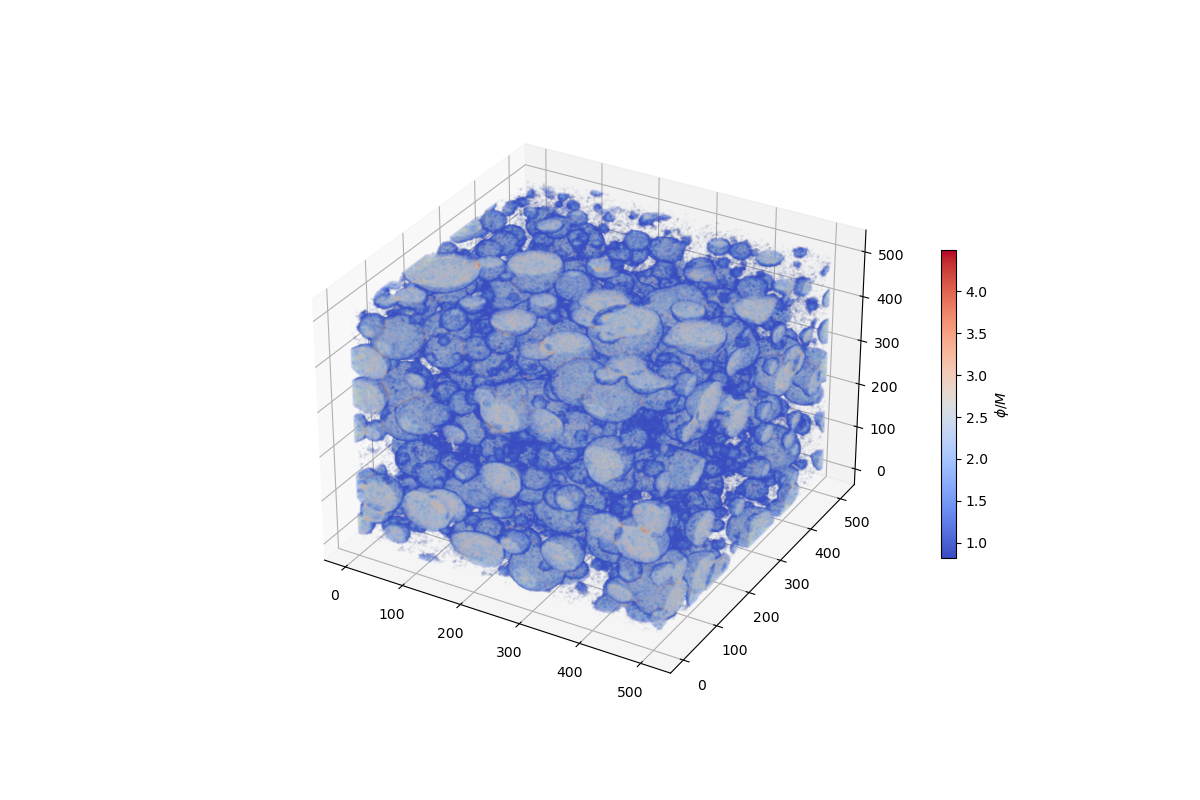}
    \caption{3D snapshot of the scalar field distribution at the times $\tilde{t} = 70.4, 105.6,140.8$ for $T_i = 1.15M$(top) and at the times $\tilde{t} = 0.0, 17.6, 35.2$ for $T_i = 1.35M$(bottom). }
    \label{fig:3d_snapshot}
\end{figure*}

In Fig.\ref{fig:3d_snapshot}, we present the three-dimensional spatial distribution of the scalar field for $T_i=1.1M, 1.35M$, where we observed the existence of oscillons and the nucleation and subsequent expansion of bubbles. Qualitatively, the number of oscillons and bubbles increases as the initial temperature rises. 

.

\end{document}